\documentclass[12pt,aps,reprint,groupedaddress,showpacs,nofootinbib,prd]{revtex4-1}
\usepackage{amsmath,amssymb,mathrsfs,graphicx,slashed,ragged2e}
\usepackage{hyperref}

\DeclareMathOperator{\csch}{csch}

\newcommand{\bsn}{\boldsymbol{\nabla}}

\begin{document}

\title{Causal properties of nonlinear gravitational waves in modified gravity}

\author{Arthur George Suvorov}
\email{suvorova@student.unimelb.edu.au}
\affiliation{School of Physics, University of Melbourne, Parkville VIC 3010, Australia}
\author{Andrew Melatos}
\email{amelatos@unimelb.edu.au}
\affiliation{School of Physics, University of Melbourne, Parkville VIC 3010, Australia}
\affiliation{Australian Research Council Centre of Excellence for Gravitational Wave Discovery (OzGrav)}


\date{\today}

\begin{abstract}

Some exact, nonlinear, vacuum gravitational wave solutions are derived for certain polynomial $f(R)$ gravities. We show that the boundaries of the gravitational domain of dependence, associated with events in polynomial $f(R)$ gravity, are not null as they are in general relativity. The implication is that electromagnetic and gravitational causality separate into distinct notions in modified gravity, which may have observable astrophysical consequences. The linear theory predicts that tachyonic instabilities occur, when the quadratic coefficient $a_{2}$ of the Taylor expansion of $f(R)$ is negative, while the exact, nonlinear, cylindrical wave solutions presented here can be superluminal for all values of $a_{2}$. Anisotropic solutions are found, whose wave-fronts trace out time- or space-like hypersurfaces with complicated geometric properties. We show that the solutions exist in $f(R)$ theories that are consistent with Solar System and pulsar timing experiments.

\end{abstract}


\maketitle

\section{Introduction}

When a field develops a localised perturbation, information about the disturbance is communicated at some finite speed to the surrounding universe. Gravitational waves (GWs) act as the energy and information transport mechanism for time-varying gravitational fields \cite{gripod,cenergy,bondi4}. Within the theory of general relativity (GR), GWs propagate in the linearised regime outside the near zone in even the most compact relativistic sources (e.g.  \cite{blanchet,owen}). Linearisation schemes are convenient because tools such as multipole expansions exist for calculating amplitudes and polarisations straightforwardly given a model of the source \cite{isaac,thorne,jetzer}. At least within GR, it is well known that the phase speed of GWs is precisely the speed of light\footnote{We adopt natural units throughout with $G=c=1$, although the constant $c$ is occasionally written explicitly for emphasis.} in both the linear and nonlinear theories \cite{einros,hawkel}. The waves propagate along null hypersurfaces in vacuum and thus the notions of (Maxwellian) electromagnetic and general relativistic causality coincide.

Fundamental inconsistencies between quantum field theories and GR suggest that a quantum theory of gravity will modify the geometry-matter relations of GR \cite{capbeyond}. Within bosonic string theories, for example, the quantization of the Polyakov action introduces scalar potentials (graviton-dilaton couplings) into the Einstein action which modifies the gravitational dynamics \cite{horo90,damour94}. Transforming into the Jordan frame shows that these dilaton-tensor theories behave like higher-order curvature theories [such as the $f(R)$ theories considered in this paper; see below], and that GR correction terms are large when the curvatures are large \cite{felice,capdilaton}. Classically speaking, therefore, GWs in string-inspired or other gravity theories may propagate differently to their GR counterparts in the vicinity of strong sources or elsewhere \cite{gasp93,roberts94}. In particular, the wave-fronts may trace out hypersurfaces, which are not null, indicating that notions of causality may differ between electromagnetic and gravitational events in modified theories of gravity. Theories with massive gravitons, for example, predict that the wave-fronts are frequency dependent, propagate slower than light, and trace out time-like hypersurfaces \cite{deser82}. A modification of the phase speed represents the simplest kind of topological adjustment that can occur in the causal structure \cite{izumi14}. Other, exotic kinds of topological structures can also occur in wave-fronts in $f(R)$ gravity. For example, there exist choices of $f$ such that the gravitational past and future of some event can have a non-empty intersection, thereby violating chronology protection \cite{hawkconj,izumi13}.

The linearised $f(R)$ theory predicts an exact dispersion relation for GWs \cite{berrygair}. However, dispersion relations in linear and nonlinear theories can have very different physical characters. Consider a scalar field theory whose equation of motion reads
\begin{equation} \label{eq:quadklein}
0 = \phi_{,tt} - \bsn^2 \phi  + V'(\phi),
\end{equation}
with scalar field $\phi$ and potential function $V$. Linearisation of equation \eqref{eq:quadklein} returns either the Klein-Gordon or massless wave equation depending on the coefficient of the linear term in $V'(\phi)$. Both the Klein-Gordon and massless wave equations admit propagating solutions with fixed propagation speeds (e.g. \cite{jeffrey}). However, depending on the form of $V$, the nonlinear dispersion relation can be modified by self-interaction \cite{kellerkog}. For example, there exist choices of $V$ such that equation \eqref{eq:quadklein} admits soliton-like solutions with arbitrary phase speeds (such as $V \propto \phi^3$), while other choices of $V$ preserve the Klein-Gordon character of the dispersion relation (such as $V \propto \cos\phi$) \cite{gonzales,gfrellis,kichen,verhulst}. Given the well-studied equivalence between $f(R)$ and scalar-tensor theories of gravity, it is reasonable to expect a similar phenomenon to occur in $f(R)$ gravity depending on the particulars of the function $f$ \cite{felice}. Hence, one must be careful when drawing conclusions about nonlinear GWs from analysis of the corresponding linearised field equations \cite{suvmel2,carr}. This phenomenon is related to the Vainshtein mechanism \cite{vainshtein}.


The purpose of this short paper is to demonstrate, by explicit construction, some topological properties of nonlinear GWs in $f(R)$ theories of gravity. We show that the predictions offered by the linear and nonlinear theories may differ significantly. In Section II we define some general notions of causality that are used throughout the paper. In Section III we present the $f(R)$ field equations and recall some results concerning phase speeds of GWs in the linear theory. In Section IV we show, by constructing two exact solutions, that these relations may fail to describe the propagation speed of nonlinear GWs, that exotic topological properties can occur in the GW-front defining the causal structure, and that the theories considered are consistent with Solar System and pulsar timing constraints. However, the analytic solutions exhibit certain artificial properties, which are likely to be avoided in more general, numerical solutions, a topic for future work.  Some brief, additional discussion regarding $f(R)$ theories and causality is presented in Section V.


\section{Causality in modified gravity}

In any physical theory where information propagates at a finite speed, a notion of causality emerges. Given an event $E_{1}$, a second event $E_{2}$ is causally connected through electromagnetic signals to $E_{1}$ provided that it lies within the null cone originating at $E_{1}$. The same two events are causally connected gravitationally, if there exists a curve joining $E_{1}$ and $E_{2}$ that is contained within the \emph{domain of dependence}, defined by the hypersurface traced out by the GW-fronts emanating from $E_{1}$ (see e.g. Hawking and Ellis \cite{hawkel} for formal definitions). In vacuum GR, the domain of dependence coincides exactly with the null cone for any event, and an unambiguous notion of causality emerges. The domain of dependence, however, depends on the structure of the field equations (since it depends on the properties of GWs) and need not coincide with the null cones in modified gravity.  Throughout this work we use the phrase `causal' to refer to gravitational causality unless otherwise stated.

\begin{figure}
\includegraphics[width=0.473\textwidth]{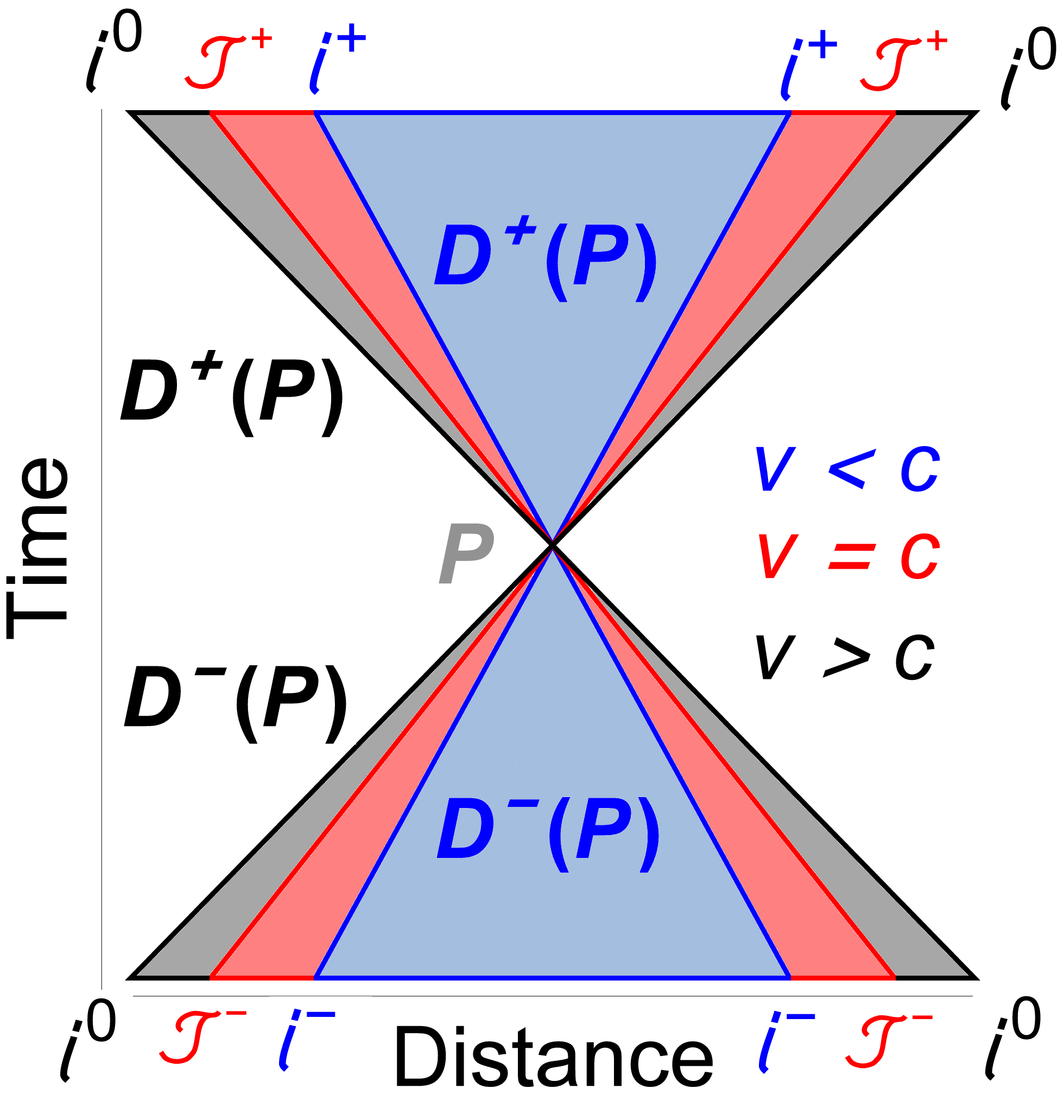}
\justifying
\caption{Cross section of the causal domains for an event $P$ for a universe where GWs propagate at speed $v$. The shaded blue region represents the causal domain of $P$, when GWs are subluminal $(v<c)$. For $v<c$, the domain $D(P) = D^{+}(P) \cup D^{-}(P)$ is strictly contained within the null cone $(v=c)$ represented by the union of the red region, which extends to future (past) null infinity $\mathcal{J}^{+}$ $(\mathcal{J}^{-})$, and the blue region, which extends to future (past) time-like infinity $i^{+}$ $(i^{-})$. The causal domain for superluminal GWs $(v>c)$ is represented by the union of the grey, red, and blue regions and extends to space-like infinity $i^{0}$. Note that, although drawn as conical structures here for simplicity, the global shape of $D(P)$ will be warped by the metric coefficients in general (see section 12.6 of \cite{choquet09}). \label{fig1GW}
}
\end{figure}

Consider a universe where GW-fronts propagate isotropically with phase speed $v$ in vacuum, and suppose some perturbation event occurs at $P$. Fig. \ref{fig1GW} illustrates three kinds of causal connection that can occur in such a universe. The domain of events which could be influenced by (influence) $P$ is known as the future (past) domain of dependence and is denoted by $D^{+}(P)$ [$D^{-}(P)$]. The set $D(P) = D^{+}(P) \cup D^{-}(P)$ represents the causal domain of the event $P$.  For $v<c$, there exist observers in Lorentz-boosted frames who see the wave travel at non-zero speeds less than $c$. The domain $D^{+}(P)$ [$D^{-}(P)$] extends to future (past) time-like infinity $i^{+}$ ($i^{-}$). In such a universe, events exist that are electromagnetically but not gravitationally connected, i.e. events which lie within the null cone originating at $P$ but not in $D(P)$. For $v=c$, gravitational events are seen at the same time as electromagnetic ones by all observers, and the domain $D^{+}(P)$ [$D^{-}(P)$] extends to future (past) null infinity $\mathcal{J}^{+}$ ($\mathcal{J}^{-}$). This is the case in vacuum GR. If GWs are superluminal with speeds $v>c$, boosted-frame observers exist who see the waves travel at arbitrarily high speeds, the domains $D^{\pm}(P)$ extend to space-like infinity $i^{0}$, and events exist that are gravitationally but not electromagnetically connected. See e.g. Refs. \cite{gonzales,bruneton} for a discussion on physical consequences.

\begin{figure}
\includegraphics[width=0.473\textwidth]{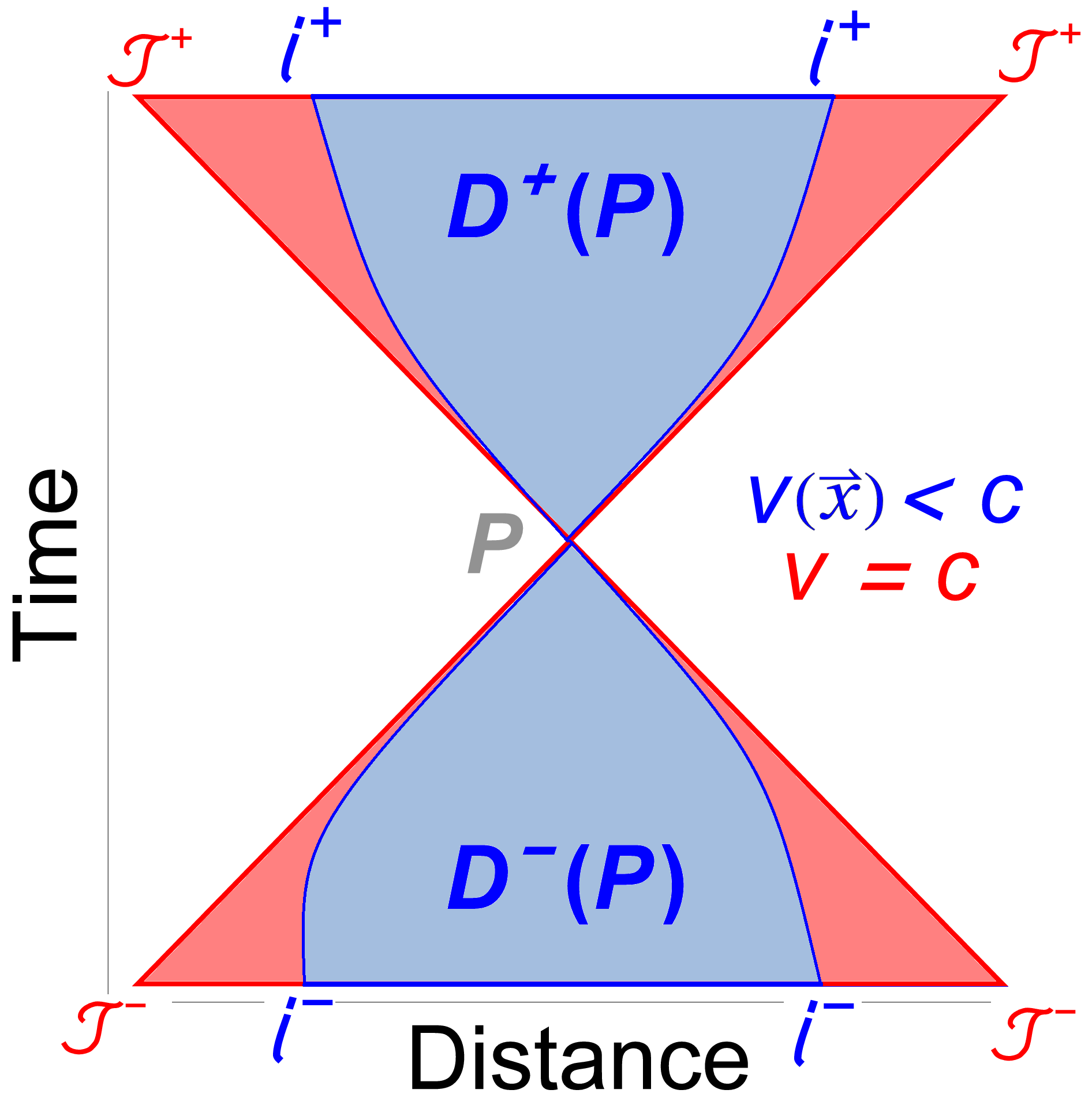}
\justifying
\caption{Cross section of the causal domain for an event $P$ within a universe where GW-fronts propagate anisotropically and subluminally. The domain $D(P) = D^{+}(P) \cup D^{-}(P)$ has some non-trivial topological structure, i.e. $\partial D(P)$ is complicated. In this particular example, $D(P)$ is completely confined within the null cone, and as such extends to future (past) time-like infinity $i^{+}$ ($i^{-}$). \label{fig2GW}
}
\end{figure}

Fig. \ref{fig2GW} represents the causal domain of the event $P$ in a universe where GW wave-fronts no longer trace out two cones (future and past) joined at $P$ but rather some other topological surface, i.e. GW propagation is not isotropic. Note that in this particular illustration we have that $D^{+}(P) \cap D^{-}(P) = \emptyset$. There is no reason to assume \emph{a priori} that this holds for general theories of gravity, i.e. closed time-like curves can exist in $D^{+}(P) \cap D^{-}(P)$ in general.\footnote{This scenario can also occur in GR for universes filled with exotic matter, e.g. the G{\"o}del solution \cite{gripod}. In this work we consider vacuum spacetimes only.}

Below we show that, in $f(R)$ gravity, both of the situations depicted in Figs. \ref{fig1GW} and \ref{fig2GW} may occur for nonlinear GWs.

\section{Linear waves}

\subsection{Field equations}

In an $f(R)$ theory of gravity, the Ricci scalar, $R$, is replaced by an arbitrary function of this quantity, $f(R)$, in the Einstein-Hilbert action. The vacuum field equations read (e.g. \cite{felice})
\begin{equation} \label{eq:fofr1}
0 = f'(R) R_{\mu \nu} -  \frac {f(R)} {2} g_{\mu \nu} + \left( g_{\mu \nu} \square - \nabla_{\mu} \nabla_{\nu} \right) f'(R),
\end{equation}
\\
where $R_{\mu \nu} = R^{\alpha}_{\mu \alpha \nu}$ is the Ricci tensor, $g_{\mu \nu}$ is the metric tensor, and $\square = \nabla_{\mu} \nabla^{\mu}$ symbolises the d'Alembert operator.


\subsection{Linear theory}
Following Berry and Gair \cite{berrygair} we consider $f$ to be an analytic function about $R=0$ so that it can be expressed as a power series,
\begin{equation} \label{eq:maclarin}
f(R) = a_{0} + a_{1} R + \frac {a_{2}} {2!} R^2 + \frac {a_{3}} {3!} R^3 + \cdots,
\end{equation}
where the $a_{i}$ are the Maclaurin coefficients \cite{capbeyond,felice}. We set $a_{0} = 0$ to expand about a Minkowski background, though some of the results carry over to other backgrounds as well, e.g. (anti-) de Sitter. Perturbing the metric according to
\begin{equation} \label{eq:metpert}
g_{\mu \nu} = \eta_{\mu \nu} + h_{\mu \nu},
\end{equation}
we find \eqref{eq:fofr1} reduces to
\begin{equation} \label{eq:linfield}
0 = \square \bar{h}_{\mu \nu}
\end{equation}
to linear order. We introduce the trace-reversed potential
\begin{equation}
\bar{h}_{\mu \nu} = a_{1} \left( h_{\mu \nu} - \frac {1} {2} h_{\sigma \rho} \eta^{\sigma \rho} \eta_{\mu \nu} \right) - a_{2} R^{(1)} \eta_{\mu \nu} ,
\end{equation}
enforce the generalised de Donder gauge $\nabla^{\mu} \bar{h}_{\mu \nu} = 0$, and write the linearised Ricci scalar [to order $\mathcal{O}(h)$] as $R^{(1)}$. Equation \eqref{eq:linfield} implies the existence of two tensor polarisation modes for general $f(R)$ theories \cite{berrygair}, just like in GR.

The trace of \eqref{eq:fofr1} shows that the linearised Ricci scalar satisfies a Klein-Gordon equation of the form
\begin{equation} \label{eq:linric}
0 = 3 a_{2} \square R^{(1)} - a_{1}  R^{(1)},
\end{equation}
indicating that there is also a propagating scalar mode for $a_{2} \neq 0$ (massive for $a_{1} \neq 0$) in addition to the two tensor modes of GR \cite{alves}. An important feature, for our purposes, is that equation \eqref{eq:linric} predicts the existence of scalar modes with group velocity\footnote{Note that there is a misplaced minus sign in equation (30) in Ref. \cite{berrygair}; Berry and Gair's no-tachyon condition should read $\Upsilon^2 < 0$ to be consistent with the usual Starobinsky $[f'(R) > 0]$ and Dolgov-Kawasaki $[f''(R) \geq 0]$ conditions \cite{starobinsky,dolgovkaw}.} $(a_{2} \neq 0)$
\begin{equation} \label{eq:groupvelocity}
c_{\textrm{g}} = \frac {\sqrt{\omega^2 - a_{1} \left(3 a_{2} \right)^{-1} }} {\omega}, 
\end{equation}
where $\omega$ is the wave frequency \cite{jeffrey,berrygair}. As such, the velocity of a linear GW in $f(R)$ gravity is uniquely determined by the value of the coefficients $a_{2}$ and $a_{1}$ given \eqref{eq:maclarin}. In particular, expression \eqref{eq:groupvelocity} demands $a_{1} a_{2} > 0$ to ensure $c_{\textrm{g}} < 1$, so that tachyonic instabilities are avoided \cite{notach,stelle}.


\section{Exact solutions}

In this section we construct three explicit examples of nonlinear wave solutions to \eqref{eq:fofr1}. In Sec. IV. A, we present a class of solutions which admit an arbitrary phase speed independent of the value of $a_{2}$, the situation depicted in Fig. \ref{fig1GW}. We also derive a class of generalised Peres waves which propagate anisotropically in Sec. IV. B, the situation depicted in Fig. \ref{fig2GW}. In both cases, we work with the function
\begin{equation} \label{eq:somef2}
f(R) = R + \frac {a_{2}} {2!} R^2 + \frac {a_{3}} {3!} R^3 + \frac {a_{k}} {\Gamma(k+1)} R^k,
\end{equation}
where $\Gamma(\xi) = \int^{\infty}_{0} d \tau \tau^{\xi-1} e^{-\tau}$, is the usual gamma function, $k$ is an integer greater than three, and at least two of the $a_{2}$, $a_{3}$, and $a_{k}$ are non-zero. 

Functions of the form \eqref{eq:somef2} have been considered in the literature as geometric models of dark energy (e.g. \cite{dark1}). In this context, the parameters appearing in \eqref{eq:somef2} have been constrained through Solar System experiments \cite{darksolar}, supernova Ia luminosity distance data \cite{darkother}, and stochastic gravitational wave background limits \cite{darksto}. While in dark energy models one typically sets $0 \leq k < 1$, the Maclaurin expansion \eqref{eq:maclarin} does not exist for $k$ in this range, since $f'(0)$ diverges and feeds into equation \eqref{eq:fofr1}. Hence, we consider $k$ to be a positive integer here. In fact the exact solutions presented below exist formally for all real $k$ (though not for any $a_{k}$), so some models may admit tachyonic gravitational waves, even when the theory cannot be linearised about a Minkowski background. A discussion of astrophysical constraints on theories of the form \eqref{eq:somef2} is presented in Sec. IV. C.

\subsection{Arbitrary phase speed}

We construct an exact solution that is cylindrically symmetric. Such solutions can be described by the Jordan-Ehlers-Kompaneets line element in Weyl coordinates $(t,\rho,\phi,z)$ \cite{jordan},
\begin{equation} \label{eq:timelineel}
ds^2 = e^{-2 \psi} \left[ e^{2 \gamma} \left( -d t^2 + d \rho^2 \right) + \rho^2 d \phi^2 \right] + e^{2\psi}  dz^{2} ,
\end{equation}
where $\psi$ and $\gamma$ are functions of $t$ and $\rho$ \cite{gripod,suvmel2}. In GR, vacuum GW solutions, represented by \eqref{eq:timelineel} or otherwise, must necessarily have unit propagation speed (see Theorem 8.8 of \cite{choquet09}). Many exact, cylindrical GW solutions are known \cite{schmidt,ashtekar2,weber,rosen}. While non-cylindrical GWs exist (e.g. in Ref. \cite{maccal} or any multipole with nonzero azimuthal wavenumber; see also below), cylindrical GWs suffice to demonstrate the points considered here.

Consider the metric \eqref{eq:timelineel} for the choices
\begin{equation} \label{eq:polypsi}
\psi = 0,
\end{equation}
and
\begin{equation} \label{eq:polygamma}
\gamma = \frac {1} {2} \ln \left\{ A \left( 1 - v^2 \right) \csch\left[ \delta + \omega \left( t - v \rho \right) \right]^{2} \right\},
\end{equation}
where $\csch(\xi) = 2/\left(e^{\xi} - e^{-\xi}\right)$ is the hyperbolic cosecant function (which is singular at $\xi = 0$), $\omega$ represents a frequency, $A \neq 0$ is an amplitude factor\footnote{Note that for $v>1$ $(v<1)$ we require $A<0$ $(A>0)$ to ensure that the metric \eqref{eq:timelineel} has a Lorentzian signature.}, $\delta$ represents a phase shift, and $v \neq 1$ is a phase velocity. In the zero-frequency limit $\omega \rightarrow 0$ we recover the Minkowski spacetime. It can be verified by direct computation that the metric given through \eqref{eq:polypsi} and \eqref{eq:polygamma} is a solution to \eqref{eq:fofr1} for $f$ given by \eqref{eq:somef2}, provided that (i) the wave satisfies an amplitude-frequency relation, as for any nonlinear wave \cite{m00k}, of the form $(a_{3} \neq 0)$
\begin{equation} \label{eq:a3param}
\frac {\omega^2} {A} =  \frac {\alpha - 3 a_{2} \left(k -2 \right)} {4 a_{3} \left(k - 3 \right)};
\end{equation}
and (ii) that the coefficient $a_{k}$ is given by
\begin{align}
a_{k} =&  \frac {2^{k-2} \Gamma(k+1) \left[ \alpha - 3 a_{2} \left( k -2 \right) \right]^{-k}} {a_{3}^{2-k} \left( k -3 \right)^{3-k} } \nonumber \\
& \times \Big[ 4 a_{3} \alpha \left( k -3 \right)   - 3 a_{2}^2 \alpha \left( k - 2 \right) \nonumber \\
& + 9 a_{2}^3 \left(k -2 \right)^{2} - 12 a_{2} a_{3} \left( k -3 \right) \left( 2 k -3 \right) \Big] \label{eq:a4param} ,
\end{align}
where 
\begin{equation} \label{eq:alphaterm}
\alpha = \left[ 9 a_{2}^2 \left( k - 2 \right)^{2} - 24 a_{3} \left( k -3 \right) \left( k - 1 \right) \right]^{1/2}.
\end{equation}

The parameter $v$, which takes any value except unity\footnote{The metric given by \eqref{eq:polypsi} and \eqref{eq:polygamma} is genuinely singular for $v=1$ since the Kretschmann invariant, $\mathcal{K}=R_{\mu \nu \alpha \beta} R^{\mu \nu \alpha \beta}$, diverges there \cite{clarkebook}.}, is the phase speed of the solitonic GW described by \eqref{eq:polypsi}--\eqref{eq:a4param}. Therefore, tachyonic GWs may exist regardless of the sign or value of $a_{2}$, contrary to the prediction \eqref{eq:groupvelocity} from the linear theory outlined in Sec III.B. Furthermore, the metric is discontinuous for $\delta = 0$ in Weyl coordinates along the curve $t = v \rho$. The causal domain for an event occurring at the origin, which emits GWs described by \eqref{eq:polypsi} and \eqref{eq:polygamma}, is represented by Fig. \ref{fig1GW} except that the case $v=c$ is not permitted. In particular, both sub- and super-luminal nonlinear modes exist regardless of the value of $a_{2}$. To the authors' knowledge, the metric given by \eqref{eq:polypsi} and \eqref{eq:polygamma} is reported here for the first time.

It should be noted that for an arbitrary value of $k$, the solution given by \eqref{eq:polypsi} and \eqref{eq:polygamma} only exists in the special case, where $a_{k}$ is given by \eqref{eq:a4param}. There is no reason \emph{a priori} to favour or disfavour theories that satisfy \eqref{eq:a4param}. The main purpose of the solution is to demonstrate that the linear criterion $a_{2} > 0$ does not guarantee the absence of tachyonic GWs. Incidentally, we show in Sec. IV. C that the constraint \eqref{eq:a4param} is consistent with various astrophysical tests for a variety of values of $a_{2}$.


As a side remark, in GR, it is well known that the gravitational collapse of stars with mass beyond the Tolman-Oppenheimer-Volkoff limit strips away information concerning the collapsing stellar remnant due to the no-hair theorems \cite{tov,nohair}. Information is removed by the formation of horizons, which causally separate regions within the spacetime \cite{information}. In $f(R)$ theories of the form \eqref{eq:somef2}, which permit the existence of superluminal GWs, it is possible that gravitational information can leak beyond the electromagnetic event horizons, which traditionally define black hole boundaries \cite{gbcaus} (cf. \cite{ori99}).

\subsection{Anisotropic propagation}

We show that arbitrary domains of dependence exist in theories of the form \eqref{eq:somef2} by considering a class of GWs that are not cylindrically symmetric. We consider a class of generalised Peres waves, which are described by the line element \cite{peres},
\begin{align} \label{eq:peres}
ds^2 =&\,\, U(\rho) \left(-dt^2 + dz^2 + d \rho^2 + \rho^2 d \phi^2 \right) \nonumber \\
&\,\,+ 2 \lambda(\rho,\phi,z+t) \left( dt + dz \right)^2
\end{align}
in Weyl coordinates for some functions $U$ and $\lambda$. In GR, the Peres waves \eqref{eq:peres} are defined with $U=1$ and represent a subclass of the well-studied pp-waves \cite{gripod}. A Peres wave represents a GW whose source is electromagnetic in origin. A perturbation of the Faraday tensor in some region of spacetime defines initial conditions, which induce GWs of the form \eqref{eq:peres} \cite{parisi07}.

In GR, the Einstein equations reduce to the requirement that $\lambda$ be harmonic in $\rho$ and $\phi$, i.e.
\begin{equation} \label{eq:harmonicl}
0 = \lambda_{,\rho \rho} + \rho^{-1} \lambda_{,\rho} + \rho^{-2} \lambda_{,\phi \phi}.
\end{equation}
However, for certain special choices of $a_{k}$ and $U$, the metric \eqref{eq:peres} is an exact solution to the field equations \eqref{eq:fofr1} for \emph{any} function $\lambda$. Explicitly, we find that the metric \eqref{eq:peres} is a solution to \eqref{eq:fofr1} for
\begin{equation} \label{eq:u1}
U(\rho) = \frac {A} {\omega^2 \rho^2 },
\end{equation}
\begin{equation} \label{eq:a41}
\frac {A} {\omega^2} = \frac {\alpha + 3 a_{2} \left( k -2 \right)} {2 \left( k -1 \right)},
\end{equation}
and
\begin{equation} \label{eq:a42}
a_{k} = \frac {2^{3-2k} 3^{1-k} \Gamma(k+1) \left( \alpha - 3 a_{2} \right) \left[ \alpha + 3 a_{2} \left( k -2 \right) \right]^{k-2} } {\left( 1 - k \right)^{k-1} \left( k - 3 \right)} ,
\end{equation}
where $A$ and $\omega$ are constants, the parameter $\alpha$ is defined through \eqref{eq:alphaterm}, $a_{2}$ and $a_{3}$ are arbitrary, and the function $\lambda$ is arbitrary. To the authors' knowledge, the metric given by \eqref{eq:peres} with \eqref{eq:u1}--\eqref{eq:a42} is reported here for the first time.

The domain of dependence associated with a generalised Peres wave is arbitrary, because the function $\lambda$ is arbitrary. Consider, for example, the case
\begin{equation}
\lambda =
\begin{cases}
\exp \left\{ - A \left[(t+z)^2 - u(\rho)^2 \right]^{-1}\right\} \textrm{ for } (t+z)^2 > u(\rho)^2, \\
0 \hspace*{4.64cm} \textrm{    otherwise}, \\
\end{cases}
\end{equation}
for some function $u$ and amplitude $A>0$. For $z=0$, the function $\lambda$ tends to zero along the curve $t^2 = u(\rho)^2$. Hence the metric \eqref{eq:peres} continuously tends to the (conformal) Minkowski spacetime outside this domain, but may have discontinuous derivatives along this boundary \cite{tomimatsufarad,belzak}. Hence the causal domain for an event occurring at the origin is defined as the region $u(\rho)^2 \leq \left(t + z \right)^2$, which is arbitrary since $u$ is arbitrary.  This situation is represented by Fig. \ref{fig2GW}, where we have $\partial D^{\pm}(O) = \{(t,\rho): u(\rho) = \pm |t|\}$ for $z=0$. Choices of $u$ exist that yield $D^{+}(O) \cap D^{-}(O) \neq \emptyset$, indicating that the notions of past and future can become conflated when generalised Peres waves with general $\lambda$ are permitted. 


\subsection{Astrophysical constraints}

In this section we review briefly, for completeness, astrophysical constraints on polynomial $f(R)$ theories given by \eqref{eq:somef2}. To this end we introduce the Parametersied Post-Newtonian (PPN) Eddington parameters $\gamma^{\text{PPN}}$ and $\beta^{\text{PPN}}$, which may be written as \cite{eddingtonparams1,eddingtonparams2}
\begin{equation} \label{eq:gammappn}
\gamma^{\text{PPN}} -1 = -\frac {f''(R)^2} {f'(R) + 2 f''(R)}, 
\end{equation}
and
\begin{equation} \label{eq:betappn}
\beta^{\text{PPN}} - 1 = \frac {f'(R) f''(R)} {8 f'(R) + 12 f''(R)^{2}} \frac {d \gamma^{\text{PPN}}} {d R},
\end{equation}
in a general $f(R)$ theory.

Table \ref{tab:ppnfr} presents a summary of data collected from recent Solar System and pulsar timing experiments when interpreted as constraints on the parameters $\gamma^{\text{PPN}}$ and $\beta^{\text{PPN}}$ \cite{ppn1,ppn2,ppn3,ppn4,ppn5}. The PPN parameters \eqref{eq:gammappn} and \eqref{eq:betappn} are evaluated at the measured value of the background scalar curvature $R_{0}$, which is determined through the Friedmann-Lema{\^{\i}}tre-Robertson-Walker relationship $R_{0} = 12 c^{-2} H_{0}^2$ where $H_{0}$ is the Hubble constant \cite{faronirev}. We assume that the Hubble radius takes the value $c H_{0}^{-1} = 4.0 \times 10^{3} \text{ Mpc}$.

\begin{table*}

\centering

  \caption{Selected Solar System and pulsar timing constraints on the PPN Eddington parameters $\gamma^{\text{PPN}}$ and $\beta^{\text{PPN}}$.}
  \begin{tabular}{lcc}
  \hline
Experiment & Constraint & Reference \\
\hline

Precession of Mercury & $-3.0 \times 10^{-3} < 2 \gamma^{\text{PPN}} - \beta^{\text{PPN}} - 1 < 3.0 \times 10^{-3}$ & \cite{ppn1} \\
Lunar Laser Ranging (Nordtvedt effect) & $-1.7 \times 10^{-3} < 4 \beta^{\text{PPN}} - \gamma^{\text{PPN}} - 3 < 0.3 \times 10^{-3} $ & \cite{ppn2} \\
Very Long Baseline Interferometry & $-4.0 \times 10^{-4} < \gamma^{\text{PPN}} -1 < 4.0 \times 10^{-4}$ & \cite{ppn3} \\
Cassini tracking & $-0.2 \times 10^{-5} < \gamma^{\text{PPN}} - 1 < 4.4 \times 10^{-5}$ & \cite{ppn4} \\
Timing of PSR B$1913$+$16$ & $ {\beta^{\text{PPN}} - 1} < 1.1 \times \left( \gamma^{\text{PPN}} -1 \right)$ & \cite{ppn5}  \\

\hline
\end{tabular}
\label{tab:ppnfr}
\end{table*}

\begin{figure}
\includegraphics[width=0.5\textwidth]{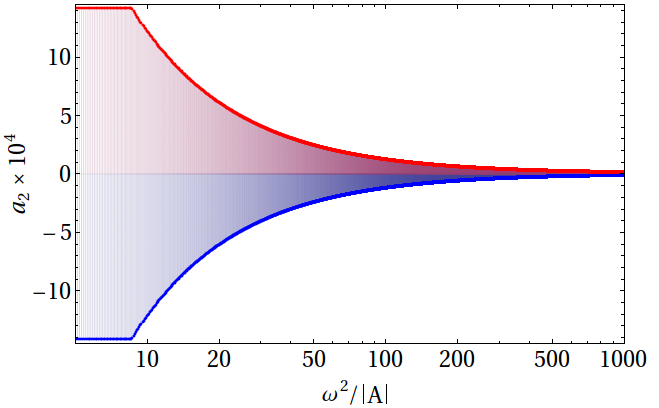}
\justifying
\caption{Allowed values of $a_{2}$ and $\omega^2/|A|$ (shaded region) in arbitrary units for the $f(R)$ theory given by \eqref{eq:somef2} consistent with the data presented in Table \ref{tab:ppnfr}, under the assumption that $a_{k}$ is given by \eqref{eq:a4param} for $k=9$. Theories for which the linear theory does (does not) predict tachyons are shaded in blue (red). \label{fig3GW}
}
\end{figure}

Using the experimental bounds on the parameters $\gamma^{\text{PPN}}$ and $\beta^{\text{PPN}}$ described in Table \ref{tab:ppnfr}, one can place constraints on the parameters $a_{2}$, $a_{3}$, and $a_{k}$ appearing in the function \eqref{eq:somef2}, and consequently constrain the set of allowed amplitude-frequency relationships \eqref{eq:a3param}. We focus on the case presented in Sec. IV. A, where we assume that the parameter $a_{k}$ is given by \eqref{eq:a4param}. In Figure \ref{fig3GW} we present values of $\omega^2 / A$ and $a_{2}$, which share the same units of length squared, consistent with the data presented in Table \ref{tab:ppnfr}, for the illustrative choice $k=9$. Figure \ref{fig3GW} demonstrates that tachyonic GWs with amplitude-frequency relation $\omega^2/A = \mu$ for some $\mu$ are permitted within astrophysically-constrained $f(R)$ theories given by \eqref{eq:somef2} provided that $a_{2}$ takes a value within the shaded region. In particular, theories for which the linear theory does (does not) predict tachyons according to \eqref{eq:groupvelocity}, are shown in the blue (red) region.


\section{Discussion}
In this paper we study the causal properties of some nonlinear GWs in vacuum $f(R)$ theories of gravity. It is found that the causal domains admit certain exotic features. The phase speeds of the waves can also be arbitrary for a wide range of functions $f(R)$, a result which does not hold in the linear regime \cite{berrygair}. The results suggest that the notion of causality is sensitive to the particulars of the modified theory of gravity under investigation \cite{gbcaus}. For example, we show that the restrictions on $f(R)$ derived previously to avoid the existence of linear superluminal GW modes \cite{berrygair,notach,stelle} must be augmented to avoid the existence of such modes in the nonlinear regime. We emphasise that it is unclear whether the exact, nonlinear solutions discussed here can actually be emitted by a realistic source with a time-dependent quadrupole moment (cf. the discussion in Ref. \cite{bondi59}). A full investigation of causality in this context relies on solving the initial boundary-value problem for a particular experiment, something falling outside the scope of this work.


Can the ideas in this paper be tested observationally? In principle, yes, although any astrophysical tests are likely to be confounded by systematic uncertainties introduced by complicated electromagnetic emission physics in the source. For example, if an event occurs which emits electromagnetic and gravitational radiation, some observers may witness the electromagnetic pulses but not the gravitational ones (if $v<c$) or vice-versa (if $v>c$). Likewise, observers at rest equidistant from the source in different planes may or may not experience the gravitational radiation if the propagation is anisotropic (see Fig. \ref{fig2GW}). Moreover, in relativistic systems like a black hole surrounded by an accretion disk, electromagnetic and gravitational wave modes induce backreactions on the disk which may not be felt simultaneously if the phase speeds are different, because of how the metric and Faraday tensors enter into the various magnetohydrodynamic couplings \cite{chandbh000,thorne741,thorne74,accretion,suvmel}. Again, designing a ``clean'' experiment of this sort is a major challenge in an astrophysical context.


The recent detection of GWs by the Laser Interferometer Gravitational-Wave Observatory (LIGO) has opened up new avenues for experimentally determining the phase speed of GWs \cite{ligo1,ligo2,conn}. Given the small number of operational interferometers at present, which makes localisation and real-time electromagnetic follow-up a difficult task, direct bounds placed on the phase speed of GWs are fairly weak at this stage \cite{ellis,collett,blas} (however see Table 5 of Ref. \cite{newgw}). To the authors' knowledge, no LIGO-related experimental bounds on the structure of $D^{\pm}$ exist. Detection of an inherently nonlinear property of GWs, such as the Christodoulou memory \cite{christo,favatam}, would be useful in this direction. It is interesting to compare predictions of the amplitude of the nonlinear memory for various theories of gravity; one can have different memory amplitudes for different GW polarisations; see e.g. equations (7) and (8) of \cite{dunish}.

\section*{Acknowledgements}
We thank the anonymous referee for their carefully considered suggestions, which improved the quality of the manuscript. This work was supported in part by an Australian Postgraduate Award, the Albert Shimmins fund, and the Australian Research Council Centre of Excellence for Gravitational Wave Discovery (OzGrav) (grant number CE170100004).




\end{document}